\documentclass[12pt]{article}
\usepackage{amsmath,amssymb}
\usepackage{graphicx,color}
\numberwithin{equation}{section}
\usepackage{cite}
\usepackage{bm}
\usepackage{yfonts}
\usepackage{dcolumn}
\newcommand{\beq}{\begin{equation}}
\newcommand{\eeq}{\end{equation}}
\newcommand{\beqa}{\begin{eqnarray}}
\newcommand{\eeqa}{\end{eqnarray}}
\newcommand{\beqar}{\begin{eqnarray*}}
\newcommand{\eeqar}{\end{eqnarray*}}
\newcommand{\wn}{\textswab{w}}
\newcommand{\bk}{{\bf{k}}}
\newcommand{\al}{\alpha}

\newcommand{\z}{\zeta}

\newcommand{\ie}{{\it i.e.,}\ }
\newcommand{\labell}[1]{\label{#1}} 
\newcommand{\reef}[1]{(\ref{#1})}

\begin{document}
\begin{titlepage}
\hfill
\vbox{
    \halign{#\hfil         \cr
          IPM/P-2008/046  \cr
           } 
      }  
\vspace*{20mm}
\begin{center}
{\Large {\bf Hydrodynamics of  ${\mathcal{N}}=6$ Superconformal Chern-Simons Theories at Strong Coupling}\\ }

\vspace*{15mm}
\vspace*{1mm}
{\large   Mohammad R. Garousi\footnote{garousi@mail.ipm.ir} and  Ahmad Ghodsi\footnote{ahmad@mail.ipm.ir}  }
\vspace*{1cm}

%


{ {Department of Physics, Ferdowsi University of Mashhad, \\ 
P.O.Box 91775-1436, Mashhad, Iran}}\\
{ and}\\
{ Institute for Research in Fundamental Sciences,\\
P.O.Box 19395-5531, Tehran, Iran}\\
\vskip 2.6 cm

\end{center}

\begin{abstract}
Using the duality conjecture between  ${\mathcal{N}}=6$ supersymmetric $U(N)_\bk\times U(N)_{-\bk}$ Chern-Simons theory and M-theory on $AdS_4\times S^7/Z_{\bk}$, we calculate the corrections to the shear viscosity  of the field theory at temperature $T$.  At strong 't Hooft coupling and at small $\bk$ level, we have considered one-loop correction to the M-theory effective action. At large $\bk$ level, we have considered  the $\alpha'$ correction to the type IIA effective action. In both cases the correction to the ratio of shear viscosity to the entropy density is positive.
\end{abstract}

\end{titlepage}

\section{Introduction}

One of the most exciting observations from AdS/CFT correspondence is the universality of the ratio of the shear viscosity $\eta$ to the entropy density $s$, for any gauge theory with an Einstein gravity dual in the limit of large color $N$ and large 't Hooft coupling $\lambda$ \cite{Policastro:2001yc,Kovtun:2003wp,Buchel:2003tz,Kovtun:2004de}. It has been conjectured in \cite{Kovtun:2004de} that this universal ratio is the lower bound of all materials at strong couplings, \ie
\beq
\frac{\eta}{s}=\frac{1}{4\pi}+\alpha
\eeq
where $\alpha$ is a positive number.

The Einstein gravity is not renormalizable, hence, it can not be considered as a consistent dual theory for a gauge theory.  The candidate theory for quantum gravity, \ie string theory/M-theory, contains Einstein gravity as well as the higher derivative corrections resulting from the stringy or loop effects. These higher derivative terms fix the correction $\alpha$ in the corresponding gauge theory. Using the AdS/CFT duality between ${\mathcal{N}}=4$ supersymmetric $SU(N)$ Yang-Mills gauge theory and type IIB string theory on $AdS_5\times S_5$, the first stringy effect to the $\eta/s$ for the ${\mathcal{N}}=4$ SYM  theory has been calculated in \cite{Benincasa:2005qc},\cite{Buchel:2004di},\cite{Myers:2008yi}. The correction is positive, consistent with the conjectured bound\footnote{See \cite{Kats:2007mq}, for a class of four-dimensional gauge theories in which the conjectured lower bound is violated.}.

Recently another example of AdS/CFT correspondence has been proposed by Aharony-Bergman-Jafferies-Maldacena (ABJM) \cite{Aharony:2008ug}. They consider
a particular brane configuration which preserves  ${\mathcal{N}}=3$ supersymmetry. At low energies, by integrating out the massive modes of the brane configuration one finds $U(N)_\bk\times U(N)_{-\bk}$ Chern-Simons conformal field theory which preserves ${\mathcal{N}}=6$ supersymmetry. This theory is renormalizable and is consistent even at high energies. By lifting the brane configuration to the M-theory they have shown that the gauge theory is equivalent to the low energy theory of $N$ coincident M2-branes in orbifold $R^8/Z_\bk$. Using the AdS/CFT correspondence, then they have  conjectured that the 3-dimensional ${\mathcal{N}}=6$ superconformal $U(N)_\bk\times U(N)_{-\bk}$ Chern-Simons-matter theory is dual to the M-theory on $AdS_4\times S^7/Z_{\bk}$. See \cite{Benna:2008zy} for recent studies in different aspects of this duality.

The 't Hooft coupling of this gauge theory is $\lambda=N/\bk$. For $\bk=1$, theory has no weak coupling regime. The viscosity  at large N has been found in \cite{Herzog:2002fn} to be $\eta=2^{3/2}\pi N^{3/2}T^2/3^3$. For $\bk>1$, however,  theory has both weak and strong coupling regimes. At strong couplings, the viscosity  becomes  $\eta=2^{3/2}\pi N^2T^2/(3^3\sqrt{\lambda})$.  On the other hand, the entropy of this theory  at the supergravity level is \cite{Klebanov:1996un,Aharony:2008ug,Garousi:2008ik}
$S=2^{7/2}\pi^2 N^{2}T^2V_2/(3^3\sqrt{\lambda})$ which gives the universal ratio $\eta/s=1/4\pi$. In this paper, we would like to examine the quantum and stringy corrections to this universal value.

The higher derivative corrections to the supergravity in general have field redefinition freedom \cite{AAT,AAT1}, so one may choose different scheme  for them. The scheme in which the  corrections are written in terms of the 11-dimensional Weyl tensor, modifies the maximal supergravity solution $AdS_4\times S^7$. On the other hand, it has been shown in \cite{Kallosh:1998qs} that the maximal solutions of supergravity are not modified by the higher derivative corrections. Hence, in this scheme the higher derivative corrections associated with  the gauge field $F_{(4)}$ influences the solution. In the scheme in which the corrections are written in terms of the 4-dimensional Weyl tensor, the maximal solution is not modified. Hence, it has been argued in \cite{Gubser:1998nz,Tseytlin:2000sf} that this scheme may include all higher derivative corrections associated with the gravity and the gauge field strength $F^{(4)}$.

An outline of this paper is as follows.
In section two we briefly review the Minkowski AdS/CFT prescription  for calculating the shear viscosity. In section 3, using the  Minkowski AdS/CFT prescription, we calculate the effect of two different schemes of one-loop correction of M-theory effective action to the shear viscosity. In section 4, using the fact that for large $\bk$ level the appropriate  description of the gauge theory is the type IIA string theory on $AdS_{4}\times CP^3$, we calculate two different schemes of  $\alpha'$ correction to the shear viscosity. In all cases,  the corrections to the classical value of the shear viscosity  is positive. Moreover, the corrections to $\eta/s$ are positive which is consistent with the $\eta/s$ bound conjecture.
\section{Shear viscosity from supergravity}
One way of  calculating the shear viscosity of a $p+1$ dimensional field theory is to use the Kubo relation
\beq
\eta=\lim_{\omega\rightarrow 0}\frac{1}{2\omega}\int dt d^px\,e^{i\omega t}\langle[T_{x_1x_2}(x),T_{x_1x_2}(0)]\rangle\,,
\eeq
which expresses the shear viscosity of a slightly non-equilibrium system in terms of the real-time correlation function of the stress energy tensor $T_{x_1x_2}$ computed in an equilibrium thermal ensemble. This relation can be written in terms of  retarded Green's function  as
\beq
\eta=\lim_{\omega\rightarrow 0}\frac{1}{2i\omega}\left[(G^{R}_{x_1x_2,x_1x_2}(\omega,0))^*-G^R_{x_1x_2,x_1x_2}(\omega,0)\right]\,,\labell{vis}
\eeq
where the  momentum space retarded Green's function is defined as
\beq
G^R_{x_1x_2,x_1x_2}(\omega,\mathbf{q})=-i\int d^{p+1}x\,e^{-iq\cdot x}\theta(t)\langle[T_{x_1x_2}(x),T_{x_1x_2}(0)]\rangle\,.
\eeq
Using the Minkowski AdS/CFT prescription \cite{Son:2002sd,Herzog:2002pc}, one relats the retarded  Green's function to the on-shell bulk action as
\beq
G^R_{x_1x_2,x_1x_2}(\omega,\mathbf{q})=\lim_{u\rightarrow 0}2{\cal F}(\omega,\mathbf{q},u)\,,\labell{GR}
\eeq
where $u=0$ is the boundary of $AdS$ on which the gauge theory lives, and ${\cal F}(\omega,\mathbf{q},u)$ is related to the on-shell action, in which the bulk metric is perturbed by the graviton which couples to the $T_{x_1x_2}$ of the gauge theory, \ie
\beq
S=S_0+\int \frac{d^{p+1}k}{(2\pi)^{p+1}}{\cal F}(\omega,\mathbf{q},u)\bigg|_{u=0}^{u=1}+\cdots \labell{Fk}
\eeq
In above equation, $S_0$ is the part of action which is independent of the perturbations, and dots refer to the terms that are cubic or higher order of perturbations. Moreover, using the equation of motion for the perturbations, the quadratic order terms in the action can be written as boundary terms, \ie at horizon of Schwarzschild $AdS$, $u=1$, and at boundary of $AdS$, $u=0$.

The thermal ${\mathcal{N}}=6$ superconformal  $U_\bk(N)\times U_{-\bk}(N)$ Chern-Simons matter theory  at strong 't Hooft coupling and for small $k$ level is described by 11-dimensional supergravity on Schwarzschild $AdS_4\times S^7/Z_\bk$. Using this gravity dual,  the shear viscosity  becomes  \cite{Herzog:2002fn}
\beq
\eta=\frac{2^{\frac32}}{3^3}\pi T^{2}\frac{(\bk N)^{\frac32}}{\bk}=\frac{2^{\frac32}}{3^3}\pi T^2\frac{N^{2}}{\sqrt{\lambda}}\labell{shear1}
\eeq
where the overall factor of $1/\bk$ is due to the orbifolding the $S^7$ and $N$ is the number of M2-branes. For large $\bk$, the orbifold circle becomes small, hence, one should reduce the M-theory to a weakly coupled type IIA theory. Using this gravity dual, one recovers again the above  shear viscosity.

\section{Quantum corrections to shear viscosity}

We have seen that the shear viscosity of the ${\mathcal{N}}=6$
superconformal  $U_\bk(N)\times U_{-\bk}(N)$ Chern-Simons matter
theory at strong 't Hooft coupling is given by \reef{shear1}. For
small $\bk$, the good description  is M-theory on $AdS_4\times
S^7/Z_\bk$. The first correction to the  result \reef{shear1} is
coming from the one-loop correction to the 11-dimensional
supergravity. We consider two different schemes for the one-loop corrections. In the next subsection, we consider the scheme in which the higher derivative terms are written in terms of the 11-dimensional Weyl tensor, and in subsection 3.2 in terms of the 4-dimensional Weyl tensor.

\subsection{Eleven-dimensional Weyl tensor}

The one-loop corrected action may be given as \cite{Russo:1997mk}
\beq
S=\frac{1}{2\kappa_{11}^2}\int d^{11}x \sqrt{-g}\bigg(R-\frac{1}{2
(4!)}F_{(4)}^2+\gamma W\bigg)\,, \labell{action} \eeq where
$\gamma=4\pi^2\kappa_{11}^{\frac43}/3$ and $W$ in terms of the
Weyl tensors is 
\beq
W=C^{hmnk}C_{pmnq}{C_h}^{rsp}{C^q}_{rsk}+\frac12
C^{hkmn}C_{pqmn}{C_h}^{rsp}{C^q}_{rsk}\,. \labell{W} 
\eeq   
In this subsection we choose the indices in Weyl tensors to run over the eleven-dimensional coordinates. 
  
The background $AdS_4\times S^7/Z_\bk$ is a solution
of the 11-dimensional supergravity which has maximal supersymmetry and should not be modified by the higher derivative corrections to this effective action\cite{Kallosh:1998qs}. In the presence of the above one-loop
correction,  the Schwarzschild $AdS_4\times S^7/Z_\bk$ solution modifies to \cite{Garousi:2008ik}
\footnote{Note that in Euclidean space, one should replace this
ansatz into the action and then find the equation of motion for
$K(u),P(u),S(u)$. However. in the Minkowski space, one has to
replace the ansatz into the Legendre transform of the action with
respect to the electric field. This is the same description that
one uses in entropy function formalism \cite{Sen:2005wa}.} \beqa
ds^2&=&S^{2n}(u)\left(\frac{r_0}{Lu}\right)^2\bigg(-K^2(u)dt^2+P^2(u)du^2+\sum_{i=1}^2dx_i^2\bigg)+L^2S^2(u)ds^2_{S^7/Z_\bk}\,,\cr
&&\cr F_{t
x_1x_2u}&=&-\frac{6r_0^4}{L^5u^4}K(u)P(u)S^{4n-7}(u)\,,\labell{ansatz1}
\eeqa where $n=-7/2$, and \beqa
K(u)&=&(1-u^3)^\frac12\left(1+\gamma X(u)\right)\,,\cr &&\cr
P(u)&=&\frac{L^2}{2r_0}(1-u^3)^{-\frac12}\left(1+\gamma
Y(u)\right)\,,\cr &&\cr S(u)&=&\left(1+\gamma
Z(u)\right)\,.\labell{perturb} \eeqa where $X(u)$ and $Y(u)$ are
\beqa
X(u)&=&\frac{1}{L^6}\left(-\frac{3568}{25}u^3-\frac{3568}{25}u^6+88u^9\right)\,,\cr
&&\cr
Y(u)&=&\frac{1}{L^6}\left(\frac{1533}{100}+\frac{3568}{25}u^3+\frac{2784}{25}u^6-536u^9\right)\,.
\eeqa 
The differential equation for $Z(u)$ does not have such a
simple solution. For extremal case, $u=0$, the tree level solution 
is modified which is not consistent with the observation made in \cite{Kallosh:1998qs}. Hence, the one-loop correction associated 
with the four-form $F_{(4)}$ which is not included in \reef{action} should affect the solution. 
It has been argued in \cite{Gubser:1998nz} that the one-loop correction associated 
with the four-form $F_{(4)}$  are scheme-dependent and 
thus are adjusted to have  the $AdS_4\times S^7/Z_\bk$ solution. 

One may  also expect that  
the one-loop correction associated with  $F_{(4)}$ modifies the differential equation for $Z(u)$ 
such that as in Schwarzschild $AdS_5\times S^5$ case, it has a simple power law solution \cite{Gubser:1998nz}. 
In \cite{Garousi:2008ik}, it has been shown that
the thermodynamic quantities such as the entropy and the
temperature do not depend on $Z(u)$. We will see that the
hydrodynamical shear viscosity quantity does not depend on $Z(u)$
either. All above quantities, however,  do depend on $X(u)$ and $Y(u)$, so they are scheme dependent.

Following the prescription for calculation the shear viscosity
\cite{Son:2002sd},\cite{Herzog:2002pc},\cite{Buchel:2004di}, one
has to perturb the metric by a graviton component that couples to
$T_{x_1x_2}$ component of the field theory at the boundary of
$AdS$, \ie \beq
ds^2\!=\!S^{2n}\left(\frac{r_0}{Lu}\right)^2\bigg[-K^2dt^2+P^2du^2+\sum_{i=1}^2dx_i^2+2\varphi
dx_1dx_2\bigg]+L^2S^2dS^2_{S^7/Z_\bk}\,, \eeq We set $L=1$. So the
perturbation is \beq
h_{x_1x_2}=\frac{r_0^2}{u^2}S^{2n}(u)\varphi(u,t,x_i)\,, \eeq We
now perform the  Fourier decomposition \beq \varphi(u,t,x_i)=\int
\frac{d^3k}{(2\pi)^3}e^{-i\omega t+ i
\mathbf{q}.\mathbf{x}}\varphi_k(u)\,, \eeq where
$k=(\omega,\mathbf{q})$. Since in the Kubo relation, the spatial
momentum is zero, we restrict our calculations to $\mathbf{q}=0$.
Replacing the perturbed metric in \reef{action} and expanding it
in terms of powers of $\varphi_k$, one finds the quadratic order
to be \beqa S_{2}&=&\frac{V_7}{2\bk
\kappa^2_{11}}\int\frac{d^3k}{(2\pi)^3}\int_0^1du\bigg\{A\varphi''_k(u)\varphi_{-k}(u)+B\varphi'_k(u)\varphi'_{-k}(u)+C\varphi'_k(u)\varphi_{-k}(u)\cr
&&\cr
&+&D\varphi_k(u)\varphi_{-k}(u)+E\varphi''_k(u)\varphi''_{-k}(u)+F\varphi''_k(u)\varphi'_{-k}(u)\bigg\}\labell{quad}\,,
\eeqa where $V_7$ is the volume of the 7-sphere. The equation of
motion for the perturbed metric is \beq
A\varphi''_k+C\varphi'_k+2D\varphi_k-\frac{d}{du}(F\varphi''_k+2B\varphi'_k+C\varphi_k)+\frac{d^2}{du^2}(2E\varphi''_k+F\varphi'_k+A\varphi_k)=0\,,\labell{eom}
\eeq and the coefficients $A,...,F$ are given in the appendix A.
In order to have a well-defined variational principle,  one must
add the following Gibbons-Hawking boundary terms to the action
\cite{Buchel:2004di}: \beq
\mathcal{K}=-A\varphi_k\varphi'_{-k}-\frac{F}{2}\varphi'_k\varphi'_{-k}+E(p_1\varphi'_k+2p_0\varphi_k)\varphi'_{-k}\,,
\eeq where $p_0,p_1$ are given by the equation of motion, \ie \beq
\varphi''_k+p_1\varphi'_k+p_0\varphi_k=O(\gamma)\,, \eeq Using the
equation of motion, the on-shell action then becomes the  boundary
term \reef{Fk} with \beqa
\mathcal{F}(\omega,0,u)&=&\frac{V_7}{2\bk
\kappa^2_{11}}\bigg[(B-A-\frac{F'}{2}+2p_0E)\varphi'_k\varphi_{-k}+\frac12(C-A')\varphi_k\varphi_{-k}+Ep_1\varphi'_k\varphi'_{-k}\cr
&&\cr
&-&E'\varphi''_k\varphi'_{-k}+E\varphi''_k\varphi'_{-k}-E\varphi'''_k\varphi_{-k}\bigg]\labell{calF}\,,
\eeqa To evaluate the above function at boundary $u=0$, one needs
to know the solution  of $\varphi_k$. After replacing the
coefficients from appendix A, the  equation of motion \reef{eom}
becomes \beqa
&&(u^7-2u^4+u)\varphi_k''+(u^6+u^3-2)\varphi_k'+u\wn^2\varphi_k
=\gamma\frac{1088}{9(-1+u^3)}\times\cr &&\cr
&\times&\!\!\!\bigg[(-1+u^3)^4\,u^3\,(u^6-\frac{14}{255}u^3+\frac{1673}{2550})\,\varphi_k''''
+24\,(-1+u^3)^3\,u^5\,(u^6-\frac{46}{85}u^3+\frac{581}{1700})\,\varphi_k'''
\cr &&\cr
&+&\!\!\!\frac{8981}{68}(-1+u^3)^2u(u^{12}\!\!-\frac{731854}{673575}u^9+\frac{18019}{44905}u^6-\frac{5871223}{71848000}u^3-\frac{456013}{30792000}+\frac{136}{8981}u^8\wn^2\cr
&&\cr
&-&\!\!\!\frac{16}{19245}u^5\wn^2+\frac{956}{96225}u^2\wn^2)\varphi_k''
+\frac{25}{34}(-1+u^3)^2(u^{12}+\frac{79471}{1875}u^9+\frac{14}{125}u^6-\frac{1409143}{400000}u^3\cr
&&\cr
&+&\!\!\!\frac{3192091}{600000}+\frac{408}{25}u^8\wn^2-\frac{56}{125}u^5\wn^2)\varphi_k'
-\frac{2809}{68}(u^{12}\!\!-\frac{323858}{210675}u^9+\frac{25233}{70225}u^6-\frac{4767127}{22472000}u^3\cr
&&\cr
&+&\!\!\!\frac{2778181}{67416000}-\frac{68}{2809}u^8\wn^2+\frac{56}{42135}u^5\wn^2-\frac{3346}{210675}u^2\wn^2)u\wn^2\varphi_k\bigg]\,,
\eeqa where $\wn=\omega/2r_0$. It is interesting to note that
while the function $Z(u)$ appears in coefficients $C$ and $D$, the
above equation of motion is independent of $Z(u)$. Hence, in order
to find the solution we do not need to know this function. To find
a solution for the above equation, one chooses the following
ansatz: \beq \varphi_k(u)=(1-u)^{\beta}G_k(u)\,, \eeq where
$G_k(u)$ should be a regular function at horizon $u=1$. We find
$G_k(u)$   perturbativly as \beq G_k(u)=G^{(0)}_k(u)+\gamma
G^{(1)}_k(u)\,, \eeq The solution for $G_k(u)$ to order $O(\wn^2)$
is regular, \ie \beq \varphi_k(u)=(1-u)^{\beta}\bigg[1+\beta
\ln(u^2+u+1)-400\beta\gamma
u^3(u^6+\frac{1796}{625}u^3+\frac{3146}{625})\bigg]\,, \eeq where
we have normalized $\varphi_k(u)$ to one at the boundary $u=0$.
The regularity condition for the whole $G_k(u)$ at supergravity
level fixes $\beta=\pm i\wn/3$. According to the Minkowski AdS/CFT
prescription, one has to choose the incoming wave at the horizon,
\ie $\beta=-i\wn/3$. As pointed out  in \cite{Buchel:2008sh}, in
order to have the regularity at  order $\gamma$,  one must check
the regularity of $O(\wn^2)$ terms at this order. By considering
$\beta=-i\wn(1+\gamma\beta_1)/3$ the coefficient of $O(\wn^2)$
terms gives the following relation \beq
\frac{1}{15}\gamma(1383+20\beta_1)r_0^3\frac{1}{u-1}+O((u-1)^0)=0\,,
\eeq which  gives $\beta=-i\wn(1-\gamma1383/20)/3$. Using the
relation between temperature and $r_0$ which has been found in
\cite{Garousi:2008ik}, \ie \beq
T=\frac{3}{2\pi}r_0(1+\gamma\frac{1383}{20})\,, \eeq One can write
the incoming wave as $(1-u)^{-i\omega/4\pi T}$. Similar relation
has been found in \cite{Buchel:2008sh} for D$_3$-brane case.

Putting the solution  into $\mathcal{F}(\omega,0,u)$ one finds the
retarded Green's function \reef{GR} to be the following value:
\beqa
G^R_{x_1x_2,x_1x_2}(\omega,0)&=&\lim_{u\rightarrow 0}\,2\mathcal{F}(\omega,0,u)\nonumber\\
&=&\lim_{u\rightarrow 0}\,\frac{V_7r_0^3}{\bk\kappa^2_{11}}\bigg[-\frac{2}{u^3}+2-i\wn(1+\frac{233308}{125}\gamma)\\ &+&\gamma(\frac{32193}{250}\frac{1}{u^3}+\frac{107167}{250}-\frac{7}{2}Z'''-\frac{7}{u}Z''-\frac{7}{u^2}Z')+O(\wn^2)\bigg]\,.\nonumber
\eeqa
The viscosity \reef{vis} then becomes
\beq
\eta=\lim_{\omega\rightarrow 0}\,\frac{V_7r_0^3\wn}{\bk\omega\kappa^2_{11}}\,(1+\frac{233308}{125}\gamma)\,.
\eeq
It is important to note that while function $Z(u)$ appears in the retarded Green's function,
 it does not appear in the shear viscosity as we have anticipated before. Using $\wn=\omega/2r_0$, $V_7=\pi^4/3$,
 $1=(\bk N)^{3/2}\kappa^2_{11}\sqrt{2}/\pi^5$, and the relation between $r_0$ and
 temperature,
one finds the viscosity in terms of the temperature to be
\beq
\eta=\frac{2^{\frac32}\pi}{3^3}T^2\frac{N^2}{\sqrt{\lambda}}(1+\frac{432041}{250}\gamma)\,.
\eeq
where the first term is the supergravity result \reef{shear1}.  Note that the correction is proportional to $\sqrt{\lambda}$, \ie $\gamma=2^{5/3}\pi^{16/3}\lambda/(3N^2)$.
The entropy density to first order of $\gamma$, has been found in \cite{Garousi:2008ik}, is
\beq
s=\frac{2^{\frac72}\pi^2}{3^3}T^2\frac{N^2}{\sqrt{\lambda}}(1+\frac{41}{250}\gamma)\,.
\eeq
The ratio of $\eta/s$ will be
\beqa
\frac{\eta}{s}&=&\frac{1}{4\pi}(1+1728\gamma)\,,\\
&=&\frac{1}{4\pi}(1+2^{\frac23}\pi^{\frac{16}{3}}1152\frac{\lambda}{N^2})\,.
\eeqa
The correction is positive which is consistent with the $\eta/s$  bound \cite{Kovtun:2004de}. Note that $\lambda/N^2=1/\bk N$ is a small number. In the next subsection we will consider another scheme for the higher derivative corrections of the supergravity. We will see that even though the temperature, entropy and the shear viscosity all change, the ratio $\eta/s$  remains the same.

\subsection{Four-dimensional Weyl tensor}

We have seen that the one-loop correction in terms of 11-dimensional Weyl tensor modifies the maximal solution $AdS_4\times S^7/Z_\bk$. This indicates that this scheme does not include all one-loop corrections associated with the 11-dimensional gravity and the gauge field strength $F_{(4)}$. In \cite{Gubser:1998nz}, it has been argued that the scheme in which the Weyl tensor is written in terms of only 4-dimensional $AdS_4$ coordinates, includes in fact the effect of gravity and the gauge field $F_{(4)}$. In this subsection, we work in this scheme, so the action is given by \reef{action} in which the Weyl tensor \reef{W} is four dimensional.

 In this scheme,    the Schwarzschild $AdS_4\times S^7/Z_\bk$ solution modifies to \reef{ansatz1} in which $n=-7/2$ and  $X(u)$, $Y(u)$, $Z(u)$ are 
\beqa
X(u)&=&\frac{1}{L^6}\left(-136 u^3-136 u^6+88 u^9\right)\,,\cr
&&\cr
Y(u)&=&\frac{1}{L^6}\left(136 u^3+136 u^6-536 u^9\right)\,, \cr 
&&\cr
Z(u)&=&\frac{1}{L^6}\frac{32}{27}\left(u^6+u^9\right)\labell{solution}\,.
\eeqa 
As we have anticipated before, the differential equation of $Z(u)$ has the above simple power law solution. 
 For extremal case, $u=0$, the tree level solution 
is not modified which is  consistent with the observation made in \cite{Kallosh:1998qs}. 

Perturbing the metric as before, one finds the following coefficients for the quadratic order action \reef{quad}:
\beqa
A&=&-\frac{r_0^3}{u^2}\bigg[4(u^3-1)+64 u^3 \gamma(33 u^9-50u^6 +17+3 u^5 \wn^2)\bigg]\,,\cr &&\cr
B&=&-\frac{r_0^3}{u^2}\bigg[3(u^3-1)+48 u^3 \gamma(39 u^9-56u^6 +17+8 u^5 \wn^2)\bigg]\,, \cr &&\cr
C&=&-\frac{r_0^3}{u^3(u^3-1)}\bigg[12(u^3-1)+\frac{64 u^6}{9}\gamma(2976u^9-5993u^6+3031u^3-14-162u^5\wn^2)\bigg]\,,\cr &&\cr
D&=&\frac{r_0^3}{u^4(u^3-1)^2}\bigg[(6u^3-6-u^2\wn^2)(u^3-1)-\frac{16 u^5}{3}\gamma(36u^{13}+460u^{10}-1000u^7+476u^4\cr &&\cr &+&28u-18u^5\wn^4-27u^9\wn^2+240u^6\wn^2-51\wn^2)\bigg]\,,\cr &&\cr
E&=&r_0^3\bigg[96 u^6 \gamma(u^3-1)^2\bigg]\,,\cr &&\cr
F&=&0\,.\nonumber
\eeqa
where $\wn=\omega/2r_0$. To evaluate the  function \reef{calF} at boundary $u=0$, one needs
to know the solution  of $\varphi_k$. After replacing the above
coefficients in the  equation of motion \reef{eom}, one finds 
\beqa
&&(u^7-2u^4+u)\varphi_k''+(u^6+u^3-2)\varphi_k'+u\wn^2\varphi_k
=\cr &&\cr
&=&\gamma\frac{96 u^3}{(-1+u^3)}\bigg[(-1+u^3)^4\,u^6\varphi_k''''
+12\,(-1+u^3)^3\,u^5\,(2u^3-1)\,\varphi_k'''
\cr &&\cr
&+&\!\!\!\frac{259}{2}(-1+u^3)^2 u (u^{9}\!\!-\frac{832}{777}u^6+\frac{60}{259}u^3-\frac{17}{777}+\frac{4}{259}u^5\wn^2)\varphi_k''
\cr &&\cr
&-&25(-1+u^3)^2(u^{9}-\frac{112}{75}u^6+\frac{17}{150}-\frac{12}{25}u^5\wn^2)\varphi_k'\cr &&\cr
&-&\frac{117}{2}(u^{9}-\frac{568}{351}u^6+\frac{20}{39}u^3-\frac{17}{351}-\frac{2}{117}u^5\wn^2)u\wn^2\varphi_k\bigg]\,,
\eeqa 
  To find
a solution for the above equation, we again  choose the following
ansatz: 
\beq 
\varphi_k(u)=(1-u)^{\beta}G_k(u)\,, 
\eeq 
where
$G_k(u)$ should be a regular function at horizon $u=1$. We find
$G_k(u)$   perturbativly as 
\beq 
G_k(u)=G^{(0)}_k(u)+\gamma
G^{(1)}_k(u)\,. 
\eeq 
The solution for $G_k(u)$ to order $O(\wn^2)$
is regular, \ie 
\beq 
\varphi_k(u)=(1-u)^{\beta}\bigg[1+\beta\ln(u^2+u+1)-400\beta\,\gamma\, u^3(u^6+\frac{71}{25}u^3+5)\bigg]\labell{phik}\,, 
\eeq 
where we have normalized $\varphi_k(u)$ to one at the boundary $u=0$.
The regularity condition for the whole $G_k(u)$ at supergravity
level fixes $\beta=\pm i\wn/3$, and  at  order $\gamma$ fixes  
$\beta=-i\wn(1+\gamma\beta_1)/3$ in which $\beta_1$ is given by the regularity of the coefficient of $O(\wn^2)$, \ie
\beq
12\gamma(80+\beta_1)r_0^3+O((u-1))=0\,,
\eeq 
which gives $\beta=-i\wn(1-80 \gamma)/3$. 

As a double check of our result, we use the fact that the incoming wave should be written as $(1-u)^{-i\omega/4\pi T}$ \cite{Buchel:2008sh}.
Using the gravity solution \reef{solution}, one  finds the temperature to be 
\beq
T=\frac{3}{2\pi}r_0(1+80 \gamma)\,, 
\eeq 
So 
the incoming wave can be written as $(1-u)^{-i\omega/4\pi T}$.

Putting the solution \reef{phik}  into $\mathcal{F}(\omega,0,u)$ one finds the
retarded Green's function \reef{GR} to be the following value:
\beqa
G^R_{x_1x_2,x_1x_2}(\omega,0)&=&\lim_{u\rightarrow 0}\,2\mathcal{F}(\omega,0,u)\nonumber\\
&=&\lim_{u\rightarrow 0}\,\frac{V_7r_0^3}{\bk\kappa^2_{11}}\bigg[-\frac{2}{u^3}+2-i\wn(1+1920\gamma) +\gamma(544-\frac{448}{9}u^3)+O(\wn^2)\bigg]\,.\nonumber
\eeqa
The viscosity \reef{vis} then becomes
\beqa
\eta&=&\lim_{\omega\rightarrow 0}\,\frac{V_7r_0^3\wn}{\bk\omega\kappa^2_{11}}\,(1+1920\gamma)\nonumber\\
&=&\frac{2^{\frac32}\pi}{3^3}T^2\frac{N^2}{\sqrt{\lambda}}(1+1760\gamma)\,.
\eeqa
where the first term is the supergravity result \reef{shear1}.

Using the gravity solution \reef{solution}, one can calculate the entropy from the Wald formula or from the free energy as in \cite{Garousi:2008ik}. The result is  
\beq
s=\frac{2^{\frac72}\pi^2}{3^3}T^2\frac{N^2}{\sqrt{\lambda}}(1+32\gamma)\,.
\eeq
The ratio of $\eta/s$ then becomes 
\beqa
\frac{\eta}{s}&=&\frac{1}{4\pi}(1+1728\gamma)\,,\\
&=&\frac{1}{4\pi}(1+2^{\frac23}\pi^{\frac{16}{3}}1152\frac{\lambda}{N^2})\,.
\eeqa
which is exactly the same value that we have found in previous section. In that section we have considered only the higher derivative terms associated with the  11-dimensional gravity, whereas, in the present section  by working with four dimensional Weyl tensor one may expect that   all higher derivative terms are include.  Hence, the above result  may indicate that the higher derivative corrections associated with the gauge field strength $F_{(4)}$ in eleven dimension  have no contribution to $\eta/s$. It would be interesting to check this explicitly. Similar calculation has been done in \cite{Myers:2008yi} for D3-brane case.

\section{Stringy corrections to shear viscosity}

The shear viscosity of the ${\mathcal{N}}=6$ superconformal  $U_{\bk}(N)\times U_{-\bk}(N)$ Chern-Simons matter theories  at strong 't Hooft coupling is given by \reef{shear1}. For large $k$, the appropriate description  is type IIA string theory on $AdS_4\times CP^3$. The  correction to the  result \reef{shear1} is coming from one-loop correction and/or $\alpha'$-correction  to the 10-dimensional supergravity.  Using the dimensional analysis, one observers that the one-loop correction to viscosity in type IIA and in M-theory has the same dependency on $N$ and $\lambda$. So, the one-loop correction should be the same as the one-loop correction in M-theory case that we have found in the previous section. In this section, we are interested in calculating the $\alpha'$-correction to the shear viscosity. We consider two different schemes for the $\alpha'$ corrections. In the next subsection, we consider the scheme in which the higher derivative terms are written in terms of the 10-dimensional Weyl tensor, and in subsection 4.2 in terms of the 4-dimensional Weyl tensor.

\subsection{Ten-dimensional Weyl tensor}

The $\alpha'$-corrected type IIA supergravity  may be given as \cite{Gross:1986iv}
\beqa
S=-\frac{1}{2\kappa^2_{10}}\int d^{10}x\,\sqrt{g}\bigg\{e^{-2\phi}(R+4(\partial\phi)^2)-\frac12\frac{1}{4!}F_{(4)}^2-\frac12\frac{1}{2!}F_{(2)}^2
+\gamma e^{-2\phi}W\bigg\}\,,\labell{eff1}
\eeqa
where $\gamma=\frac18\zeta(3)(\al')^3$.  
The indices in the Weyl tensors run over the ten dimensional coordinates. So the scheme here is similar to the one considered in section 3.1. 

The background Schwarzschild $AdS_4\times CP^3$ is a solution of the 10-dimensional type IIA supergravity. In the presence of the above $\alpha'$ corrections,  the solution modifies to \cite{Garousi:2008ik}
\beqa
ds^2&=&\frac{L}{\bk R_{11}}\bigg(S^{-6}(u)\left(\frac{r_0}{Lu}\right)^2(-K^2(u)dt^2+P^2(u)du^2+\sum_{i=1}^2dx_i^2)+L^2S^2(u)ds^2_{CP^3}\bigg)\,,\cr &&\cr
F_{t x_1x_2u}&=&-\frac{6r_0^4}{L^5u^4}K(u)P(u)S^{-18}(u)\,,\cr &&\cr
F_{(2)}&=&R_{11}\bk d\omega\,,\labell{ansatz}
\eeqa
where $R_{11}=g_s^{2/3} l_p$ is the radius of the eleventh direction, and
\beqa
\omega&=&\frac{1}{2}(\cos^2\xi-\sin^2\xi)d\psi +\frac{1}{2}\cos^2\xi
\cos\theta_1 d\phi_1+ \frac{1}{2}\sin^2\xi \cos\theta_2 d\phi_2 \nonumber\\
ds^2_{CP^3}&=&d\xi^2+\cos\xi^2\sin^2\xi(d\psi+\frac{\cos\theta_1}{2}d\phi_1-
\frac{\cos\theta_2}{2}d\phi_2)^2\cr &&\cr
&+&\frac{1}{4}\cos^2\xi(d\theta_1^2+\sin^2\theta_1
d\phi_1^2)+\frac{1}{4}\sin^2\xi(d\theta_2^2+\sin^2\theta_2
d\phi_2^2)\,.
\eeqa
and $K(u),\,P(u),\,S(u)$ are those appear in \reef{perturb} in which $X(u)$ and $Y(u)$ are
\beqa
X(u)&=&\frac{1}{L^6}(\frac{\bk R_{11}}{L})^3\left(-136u^3-136u^6+88u^9\right)\nonumber\\
Y(u)&=&\frac{1}{L^6}(\frac{\bk
R_{11}}{L})^3\left(4+136u^3+136u^6-536u^9\right)\,. \eeqa The
differential equation for $Z(u)$  does not have  a power law
solution. Moreover, for the extremal case, $u=0$, the supergravity solution is modified. This indicates that  the  higher derivative corrections associated with the gauge fields $F^{(4)}$ and $F^{(2)}$ which are not included in action \reef{eff1} affects the solution. 

To find the shear viscosity, one has to perturb the metric
as \beqa ds^2&=&ds^2|_{\rm unperturb}+\frac{L}{\bk
R_{11}}S^{-6}(u)\frac{r_0^2}{L^2u^2}\, 2\,\varphi\, dx_1dx_2\labell{pert}\,.
\eeqa By Fourier expanding the perturbation as before, one finds
the following expression for $\mathcal{F}(\omega, 0, u)$: \beqa
\mathcal{F}(\omega,0,u)&=&\frac{V_{CP^3}}{2\kappa^2_{10}}\bigg[(B-A-\frac{F'}{2}+2p_0E)\varphi'_k\varphi_{-k}+\frac12(C-A')\varphi_k\varphi_{-k}+Ep_1\varphi'_k\varphi'_{-k}\cr
&&\cr
&-&E'\varphi''_k\varphi'_{-k}+E\varphi''_k\varphi'_{-k}-E\varphi'''_k\varphi_{-k}\bigg]\labell{calF'}\,,
\eeqa where the coefficients $A,...,F$ are given in the appendix B.
Replacing them into \reef{eom}, one finds the following equation
of motion \beqa
&&(u^7-2u^4+u)\varphi_k''+(u^6+u^3-2)\varphi_k'+u\wn^2\varphi_k
=\gamma\frac{k^3R_{11}^3}{L^9}\frac{120}{u^3-1}\times\cr &&\cr
&\times
&\bigg[(u^3-1)^4\,(u^6+\frac35)\,u^3\,\varphi_k''''+24\,(u^3-1)^3\,(u^6-\frac12u^3+\frac{3}{10})\,u^5\,\varphi_k'''
\cr &&\cr
&+&132(u^3-1)^2(u^{12}-\frac{521}{495}u^9+\frac{119}{330}u^6-\frac{299}{3960}u^3-\frac{19}{1320}+\frac{1}{66}\wn^2u^8+\frac{1}{110}\wn^2u^2)u\,\varphi_k''\cr
&&\cr
&+&\frac{532}{15}\,(u^3-1)^2\,(u^9-\frac{6}{133}u^6-\frac{83}{1064}u^3+\frac{3}{28}+\frac{45}{133}\wn^2u^8)\,\varphi_k'\cr
&&\cr
&-&\frac{209}{5}\,u\wn^2\,(u^{12}-\frac{88}{57}u^9+\frac{7}{19}u^6-\frac{89}{418}u^3+\frac{5}{114}-\frac{5}{209}\wn^2u^8-\frac{3}{209}\wn^2u^2)\,\varphi_k\bigg]\,,
\eeqa where $\wn=\omega\frac{L^2}{2r_0}$. We again solve the
equation of motion perturbativly and find the following result
\beq \varphi_k(u)=\frac{L^3}{\bk R_{11}}(1-u)^{\beta}\bigg[1+\beta
\ln(u^2+u+1)-400\beta\gamma\frac{\bk^3R_{11}^3}{L^9}
u^3(u^6+\frac{71}{25}u^3+5)\bigg]\,, \eeq In this case the
regularity condition at supergravity level gives $\beta=-i\wn/3$,
and   regularity of $\gamma$ order  terms computes the correction
to $\beta$. If we consider
$\beta=-i\wn(1+\beta_1\gamma{\bk^3R_{11}^3}/{L^9})/3$ then \beq
\frac{1}{12}r_0^3\gamma(76+\beta_1)\frac{1}{(u-1)}+O((u-1)^0)=0\,,
\eeq which gives rise to
$\beta=-i\wn(1-76\gamma{\bk^3R_{11}^3}/{L^9})/3$.  Putting the
solution into the $\mathcal{F}(\omega,0,u)$ one finds the
following value for the retarded Green's function: \beqa
G^R_{x_1x_2,x_1x_2}(\omega,0)&=&\lim_{u\rightarrow 0}\,2\mathcal{F}(\omega,0,u)\nonumber\\
&=&\lim_{u\rightarrow 0}\,\frac{V_{CP^3}r_0^3}{\kappa^2_{10}}\frac{16L^3}{\bk R_{11}}\bigg[-\frac{1}{8u^3}+\frac18-\frac{1}{16}i\wn(1+1840\gamma\frac{\bk^3 R_{11}^3}{L^9})\\ &+&\gamma\frac{\bk^3 R_{11}^3}{L^9}\,(\frac{21}{2}\frac{1}{u^3}+\frac{47}{2}-\frac{3}{16}Z'''-\frac{3}{8u}Z''-\frac{3}{8u^2}Z')+O(\wn^2)\bigg]\,,\nonumber
\eeqa
The shear viscosity \reef{vis} then becomes
\beq
\eta=\lim_{\omega\rightarrow 0}\,\frac{V_{CP^3}r_0^3\wn}{\omega\kappa^2_{10}}\frac{L^3}{\bk R_{11}}(1+1840\gamma\frac{\bk^3R_{11}^3}{L^9})\,.
\eeq
The temperature in terms of $r_0$ is \cite{Garousi:2008ik}
\beq
T=\frac{3r_0}{2\pi L^2}\left(1+76\frac{\gamma}{L^6}\left(\frac{\bk R_{11}}{L}\right)^3\right)\,,
\eeq
The shear viscosity in terms of temperature becomes
\beq
\eta=\frac{2\pi^2L^9V_{CP^3}T^2}{9\bk R_{11}\kappa_{10}^2}\bigg(1+1688\frac{\gamma}{L^6}\left(\frac{\bk R_{11}}{L}\right)^3\bigg)\,.
\eeq
Using the relations $\frac{2\pi R_{11}}{ \kappa_{11}^2}=\frac{1}{\kappa_{10}^2}$,  $2\pi$ Vol$(CP^3)=$ Vol$(S^7)$, $l_{11}=g_2^{1/3} l_s$ and  $2\kappa^2_{11}=(2\pi l_{11})^9/2\pi$,     one can write the viscosity in terms of 't Hooft coupling and the number of colors as
\beq
\eta=\frac{2^{\frac32}\pi}{3^3}\frac{N^2}{\sqrt{\lambda}}T^2\left(1+\frac{1688\z(3)}{\pi^{3}2^{\frac{23}{2}}\lambda^{3/2}}\right)\,.
\eeq
The first term is the supergravity result and the second term is its $\alpha'$ correction. The entropy density to first order of $\gamma$ has been found in \cite{Garousi:2008ik} to be
\beq
s=\frac{2^{\frac72}\pi^2}{3^3}\frac{N^2}{\sqrt{\lambda}}T^2\left\{1-\frac{5\z(3)}{\pi^{3}2^{\frac{17}{2}}\lambda^{3/2}}\right\}\,.
\eeq
The ratio of $\eta/s$ in this case is
\beq
\frac{\eta}{s}=\frac{1}{4\pi}\left(1+27\frac{\z(3)}{\pi^{3}2^{\frac{11}{2}}\lambda^{3/2}}\right)\,.\labell{etas2}
\eeq
Again the correction is positive which is consistent with the $\eta/s$  bound \cite{Kovtun:2004de}. The above results do not include the higher derivative terms associated with the gauge fields $F_{(4)}$ and $F_{(2)}$. In the next subsection, we consider another scheme which may include the important part of these higher derivative terms  that one needs for calculating the temperature, entropy and the shear viscosity..

\subsection{Four-dimensional Weyl tensor}

In this subsection, we consider the scheme in which the higher derivative terms are given by action \reef{eff1} in which the indices on the Weyl tensor run over the 4-dimensional $AdS_4$ part of the 10-dimensional space. 
Using the ansatz \reef{ansatz}, the solution for 
 $X(u)$ and $Y(u)$ are
\beqa
X(u)&=&\frac{1}{L^6}(\frac{\bk R_{11}}{L})^3\left(-136u^3-136u^6+88u^9\right)\nonumber\\
Y(u)&=&\frac{1}{L^6}(\frac{\bk
R_{11}}{L})^3\left(136u^3+136u^6-536u^9\right)\,.\labell{solution2} \eeqa However, the
differential equation for $Z(u)$, \ie  
\beq
2u^2(u^3-1)Z''(u)+2u(u^3+2)Z'(u)+35Z(u)-216 \frac{1}{L^6}(\frac{\bk R_{11}}{L})^3 u^{12}=0\,.
\eeq
does not have  such a simple power law
solution. For extremal case , $u=0$, the solution for $X$ and $Y$ does not modify the supergravity solution. This may indicate that this scheme includes the effect of gravity as well as the gauge fields. On the other hand, one may  expect that when all higher derivative terms are included, the differential equation of $Z(u)$  should then have simple power law solution as in $AdS_5\times S^5$ and $AdS_4\times S^7$. One may conclude  this scheme  does not include all the higher derivative terms. However, those higher derivative terms that are not included in this scheme may  modify only the differential equation of $Z(u)$ to have power law solution. Since the quantities $T,s,\eta$ are independent of $Z$, one may  expect that this scheme produces correctly these quantities.

To find the shear viscosity, one has to perturb the above metric 
as in \reef{pert}. Doing the same steps as before, one finds the following coefficient for the function \reef{calF'}:  
\beqa
A&=&-\frac{L^3}{\bk R_{11}}\frac{r_0^3}{u^2}\bigg[\frac{(u^3-1)}{4}+4 u^3\gamma\frac{\bk^3R_{11}^3}{L^9}(33 u^9-50u^6 +17+3 u^5 \wn^2)\bigg]\,,\cr &&\cr
B&=&-\frac{L^3}{\bk R_{11}}\frac{3r_0^3}{u^2}\bigg[\frac{(u^3-1)}{16}+u^3\gamma\frac{\bk^3R_{11}^3}{L^9}(39 u^9-56u^6 +17+8 u^5 \wn^2)\bigg]\,, \cr &&\cr
C&=&-\frac34\frac{L^3}{\bk R_{11}}\frac{r_0^3}{u^3(u^3-1)}\bigg[(u^3-1)+u\gamma\frac{\bk^3R_{11}^3}{L^9}(1776u^{14}-3568u^{11}+1792u^8\cr && \cr
&-&96\wn^2u^{10}-(u^3-1)^2Z')\bigg]\,,\cr &&\cr
D&=&\frac{1}{16}\frac{L^3}{\bk R_{11}}\frac{r_0^3}{u^4(u^3-1)^2}\bigg[(6u^3-6-u^2\wn^2)(u^3-1)
-2u\gamma\frac{\bk^3R_{11}^3}{L^9}(432u^{17}+480u^{14}\cr &&\cr 
&-&2256u^{11}+1344u^8
-48\wn^4u^{9}-72\wn^2u^{13}+640\wn^2u^{10}-136\wn^2u^4\cr &&\cr 
&+&3u(u^3-1)^3 Z''+3(u^{3}-1)^2Z')\bigg]\,,\cr &&\cr
E&=&6u^6r_0^3\gamma\frac{\bk^2R_{11}^2}{L^6}(u^3-1)^2\,,\cr &&\cr
F&=&0\,.\nonumber
\eeqa
where $\wn=\omega\frac{L^2}{2r_0}$. Replacing them into \reef{eom}, one finds the following equation
of motion 
\beqa
&&(u^7-2u^4+u)\varphi_k''+(u^6+u^3-2)\varphi_k'+u\wn^2\varphi_k
=\cr &&\cr
&=&\gamma\frac{k^3R_{11}^3}{L^9}\frac{96 u^3}{(-1+u^3)}\bigg[(-1+u^3)^4\,u^6\varphi_k''''
+12\,(-1+u^3)^3\,u^5\,(2u^3-1)\,\varphi_k'''
\cr &&\cr
&+&\!\!\!\frac{259}{2}(-1+u^3)^2 u (u^{9}\!\!-\frac{832}{777}u^6+\frac{60}{259}u^3-\frac{17}{777}+\frac{4}{259}u^5\wn^2)\varphi_k''
\cr &&\cr
&-&25(-1+u^3)^2(u^{9}-\frac{112}{75}u^6+\frac{17}{150}-\frac{12}{25}u^5\wn^2)\varphi_k'\cr &&\cr
&-&\frac{117}{2}(u^{9}-\frac{568}{351}u^6+\frac{20}{39}u^3-\frac{17}{351}-\frac{2}{117}u^5\wn^2)u\wn^2\varphi_k\bigg]\,,
\eeqa 
 We again solve the
equation of motion perturbativly and find the following result
\beq \varphi_k(u)=\frac{L^3}{\bk R_{11}}(1-u)^{\beta}\bigg[1+\beta
\ln(u^2+u+1)-400\beta\gamma\frac{\bk^3R_{11}^3}{L^9}
u^3(u^6+\frac{71}{25}u^3+5)\bigg]\,, \eeq In this case the
regularity condition at supergravity level gives $\beta=-i\wn/3$,
and   regularity of $\gamma$ order  terms computes the correction
to $\beta$. If we consider
$\beta=-i\wn(1+\beta_1\gamma{\bk^3R_{11}^3}/{L^9})/3$ then 
\beq
\frac{3}{4}r_0^3\gamma(80+\beta_1)+O((u-1))=0\,,
\eeq 
which gives rise to
$\beta=-i\wn(1-80\gamma{\bk^3R_{11}^3}/{L^9})/3$.  Putting the
solution into the $\mathcal{F}(\omega,0,u)$ one finds the
following value for the retarded Green's function: \beqa
G^R_{x_1x_2,x_1x_2}(\omega,0)&=&\lim_{u\rightarrow 0}\,2\mathcal{F}(\omega,0,u)\nonumber\\
&=&\lim_{u\rightarrow 0}\,\frac{V_{CP^3}r_0^3}{\kappa^2_{10}}\frac{16L^3}{\bk R_{11}}\bigg[\frac{1}{8u^3}-\frac18-\frac{1}{16}i\wn(1+1920\gamma\frac{\bk^3 R_{11}^3}{L^9})\\ &+&\frac{1}{16}\gamma\frac{\bk^3 R_{11}^3}{L^9}\,(\frac{544}{u^3}-3 Z'''-\frac{6}{u}Z''-\frac{6}{u^2}Z')+O(\wn^2)\bigg]\,,\nonumber
\eeqa
The shear viscosity \reef{vis} then becomes
\beq
\eta=\lim_{\omega\rightarrow 0}\,\frac{V_{CP^3}r_0^3\wn}{\omega\kappa^2_{10}}\frac{L^3}{\bk R_{11}}(1+1920\gamma\frac{\bk^3R_{11}^3}{L^9})\,.
\eeq

Using the solution \reef{solution2}, one finds the temperature to be 
\beq
T=\frac{3r_0}{2\pi L^2}\left(1+80\frac{\gamma}{L^6}\left(\frac{\bk R_{11}}{L}\right)^3\right)\,,
\eeq
The shear viscosity in terms of temperature then becomes
\beqa
\eta&=&\frac{2\pi^2L^9V_{CP^3}T^2}{9\bk R_{11}\kappa_{10}^2}\bigg(1+1760\frac{\gamma}{L^6}\left(\frac{\bk R_{11}}{L}\right)^3\bigg)\nonumber\\
&=&
\frac{2^{\frac32}\pi}{3^3}\frac{N^2}{\sqrt{\lambda}}T^2\left(1+\frac{1760\z(3)}{\pi^{3}2^{\frac{23}{2}}\lambda^{3/2}}\right)\,.
\eeqa
The first term is the supergravity result and the second term is its $\alpha'$ correction. 

Using the solution \reef{solution2}, one finds the entropy to the first order of $\gamma$ to be 
\beq
s=\frac{2^{\frac72}\pi^2}{3^3}\frac{N^2}{\sqrt{\lambda}}T^2\left\{1+\frac{32\z(3)}{\pi^{3}2^{\frac{17}{2}}\lambda^{3/2}}\right\}\,.
\eeq
The ratio of $\eta/s$  is then
\beq
\frac{\eta}{s}=\frac{1}{4\pi}\left(1+1728\frac{\z(3)}{\pi^{3}2^{\frac{11}{2}}\lambda^{3/2}}\right)\,.
\eeq
 We expect that the above result includes effect of all higher derivative terms. The above result is not the same as the result in \reef{etas2} which includes only the 10-dimensional gravity effects. On the other hand, as we have argued before in section 3.2, we expect that the higher derivative terms associated with the gauge field $F_{(4)}$ have no contribution to $\eta/s$. Hence, the higher derivative terms associated with the gauge field strength $F_{(2)}$ in ten dimension should have contribution to $\eta/s$ in \reef{etas2}. It would be interesting to perform these calculations explicitly along the line of \cite{Myers:2008yi}.

\section*{Acknowledgment}
A. G. would like to thank K. Bitaghsir for discussion.
\section*{A. Coefficients in M-theory}
The coefficients of equation of motion \reef{eom} are
\beqa
A&=&-\frac{r_0^3}{u^2}\bigg[4(u^3-1)+\frac{\gamma}{3375}(7128000u^{12}-10437120u^9-453600u^6+2893509u^3\cr && \cr &+&869211+480000u^8\wn^2+44800u^5\wn^2-626080u^2\wn^2)\bigg]\,,\cr &&\cr
B&=&-\frac{r_0^3}{u^2}\bigg[3(u^3-1)+\frac{\gamma}{13500}(24936000u^{12}-36453760u^9-3780000u^6+4665327u^3\cr && \cr
&+&5184000u^8\wn^2-362880u^2\wn^2-2038127)\bigg]\,, \cr &&\cr
C&=&-\frac{r_0^3}{u^3(u^3-1)}\bigg[12(u^3-1)+\frac{\gamma}{3375}(71928000u^{15}-143052480u^{12}+70489440u^9\cr && \cr &+&635040u^6-2607633u^3+2607633-4056000u^{11}\wn^2-291200u^8\wn^2-218960u^5\wn^2
\cr && \cr &-&1161440u^2\wn^2-47250Z'u^7+94500Z'u^4-47250Z'u)\bigg]\,,\cr &&\cr
D&=&\frac{r_0^3}{u^4(u^3-1)^2}\bigg[(6u^3-6-u^2\wn^2)(u^3-1)-\frac{\gamma}{13500}(11664000u^{18}-103680u^{15}\cr &&\cr
&-&27708480u^{12}+9072000u^9+12291426u^6-10430532u^3+5215266-1632000u^{10}\wn^4\cr &&\cr
&+&89600u^7\wn^4-1070720u^4\wn^4-1944000u^{14}\wn^2+17159040u^{11}\wn^2-151200u^8\wn^2\cr &&\cr &-&4671621u^5\wn^2+455301u^2\wn^2 -94500u^{11}Z''+283500u^8Z''-283500u^5Z''\cr &&\cr
&+&94500u^2Z''-94500Z'u^{10}+283500Z'u^4-189000Z'u)\bigg]\,,\cr &&\cr
E&=&r_0^3\bigg[\frac{32}{675}\gamma(u^3-1)^2(2550u^6-140u^3+1673)\bigg]\,,\cr &&\cr
F&=&r_0^3\bigg[\frac{448}{9u}\gamma(u^3-1)(u^3+2)(u^6-\frac{4}{15}u^3+\frac{1037}{300})\bigg]\,.\nonumber
\eeqa
\newpage
\section*{B. Coefficients in type IIA theory}
The coefficients are
\beqa
A&=&-\frac{16L^3}{\bk R_{11}}\frac{r_0^3}{u^2}\bigg[\frac{(u^3-1)}{4}+\gamma\frac{\bk^3R_{11}^3}{L^9}(132u^{12}-200u^9+47u^3+21+9\wn^2u^8-9\wn^2u^2)\bigg]\,,\cr &&\cr
B&=&-\frac{4L^3}{\bk R_{11}}\frac{3r_0^3}{u^2}\bigg[\frac{(u^3-1)}{4}+\gamma\frac{\bk^3R_{11}^3}{L^9}(154u^{12}-232u^9-14u^6+23u^3-3+32\wn^2u^8)\bigg]\,, \cr &&\cr
C&=&-\frac{4L^3}{\bk R_{11}}\frac{3r_0^3}{u^3(u^3-1)}\bigg[(u^3-1)+\gamma\frac{\bk^3R_{11}^3}{L^9}(1776u^{15}-3568u^{12}+1792u^9-84u^3+84\cr && \cr
&-&100\wn^2u^{11}-8\wn^2u^8-12\wn^2u^5-24\wn^2u^2-u^7Z'+2u^4Z'-uZ')\bigg]\,,\cr &&\cr
D&=&\frac{L^3}{\bk R_{11}}\frac{r_0^3}{u^4(u^3-1)^2}\bigg[(6u^3-6-u^2\wn^2)(u^3-1)\cr &&\cr
&-&2\gamma\frac{\bk^3R_{11}^3}{L^9}(432u^{18}+480u^{15}-2256u^{12}+1344u^9+252u^6-504u^3+252\cr &&\cr
&-&60\wn^4u^{10}-36\wn^4u^4-72\wn^2u^{14}+640\wn^2u^{11}-174\wn^2u^5+38\wn^2u^2\cr &&\cr &-&3u^{11}Z''+9u^8Z''-9u^5Z''+3u^2Z''-3u^{10}Z'+9u^4Z'-6uZ')\bigg]\,,\cr &&\cr
E&=&16r_0^3\bigg[\frac{3}{2}\gamma\frac{\bk^2R_{11}^2}{L^6}(u^3-1)^2(5u^6+3)\bigg]\,,\cr &&\cr
F&=&16r_0^3\bigg[\frac{3}{u}\gamma\frac{\bk^2R_{11}^2}{L^6}(u^3-1)(u^3+2)(u^6+3)\bigg]\,.\nonumber
\eeqa


\begin{thebibliography}{99}
\bibitem{Policastro:2001yc}
  G.~Policastro, D.~T.~Son and A.~O.~Starinets,
  Phys.\ Rev.\ Lett.\  {\bf 87}, 081601 (2001)
  [arXiv:hep-th/0104066].
\bibitem{Kovtun:2003wp}
  P.~Kovtun, D.~T.~Son and A.~O.~Starinets,
  JHEP {\bf 0310}, 064 (2003)
  [arXiv:hep-th/0309213].
\bibitem{Buchel:2003tz}
  A.~Buchel and J.~T.~Liu,
  Phys.\ Rev.\ Lett.\  {\bf 93}, 090602 (2004)
  [arXiv:hep-th/0311175].
\bibitem{Kovtun:2004de}
  P.~Kovtun, D.~T.~Son and A.~O.~Starinets,
  Phys.\ Rev.\ Lett.\  {\bf 94}, 111601 (2005)
  [arXiv:hep-th/0405231].
\bibitem{Benincasa:2005qc}
  P.~Benincasa and A.~Buchel,
  JHEP {\bf 0601}, 103 (2006)
  [arXiv:hep-th/0510041].
\bibitem{Buchel:2004di}
  A.~Buchel, J.~T.~Liu and A.~O.~Starinets,
  Nucl.\ Phys.\  B {\bf 707}, 56 (2005)
  [arXiv:hep-th/0406264].
\bibitem{Myers:2008yi}
  R.~C.~Myers, M.~F.~Paulos and A.~Sinha,
  arXiv:0806.2156 [hep-th].
\bibitem{Kats:2007mq}
  Y.~Kats and P.~Petrov,
  arXiv:0712.0743 [hep-th];
  M.~Brigante, H.~Liu, R.~C.~Myers, S.~Shenker and S.~Yaida,
  Phys.\ Rev.\  D {\bf 77}, 126006 (2008)
  [arXiv:0712.0805 [hep-th]];
  M.~Brigante, H.~Liu, R.~C.~Myers, S.~Shenker and S.~Yaida,
  Phys.\ Rev.\ Lett.\  {\bf 100}, 191601 (2008)
  [arXiv:0802.3318 [hep-th]].








\bibitem{Aharony:2008ug}
  O.~Aharony, O.~Bergman, D.~L.~Jafferis and J.~Maldacena,
  arXiv:0806.1218 .



\bibitem{Benna:2008zy}
  M.~Benna, I.~Klebanov, T.~Klose and M.~Smedback,
  arXiv:0806.1519 .
  M.~Blau and M.~O'Loughlin,
  arXiv:0806.3253 .
  T.~Nishioka and T.~Takayanagi,
  arXiv:0806.3391 .
  Y.~Honma, S.~Iso, Y.~Sumitomo and S.~Zhang,
   ``Scaling limit of N=6 superconformal Chern-Simons theories and Lorentzian
  arXiv:0806.3498 .
  Y.~Imamura and K.~Kimura,
  arXiv:0806.3727 .
  J.~A.~Minahan and K.~Zarembo,
  arXiv:0806.3951 .
  D.~Gaiotto, S.~Giombi and X.~Yin,
  arXiv:0806.4589 .
  C.~Ahn,
  arXiv:0806.4807 .
  G.~Arutyunov and S.~Frolov,
  arXiv:0806.4940 .
  B.~j.~Stefanski,
  arXiv:0806.4948 .
  G.~Grignani, T.~Harmark and M.~Orselli,
  arXiv:0806.4959 .
  K.~Hosomichi, K.~M.~Lee, S.~Lee, S.~Lee and J.~Park,
  arXiv:0806.4977 .
  K.~Okuyama,
  arXiv:0807.0047 .
  J.~Bagger and N.~Lambert,
  arXiv:0807.0163 .
  S.~Terashima,
  arXiv:0807.0197 .
  G.~Grignani, T.~Harmark, M.~Orselli and G.~W.~Semenoff,
  arXiv:0807.0205 .
  S.~Terashima and F.~Yagi,
  arXiv:0807.0368 .
  N.~Gromov and P.~Vieira,
  arXiv:0807.0437 .
  C.~Ahn and P.~Bozhilov,
  arXiv:0807.0566 .
  N.~Gromov and P.~Vieira,
  arXiv:0807.0777 .
  B.~Chen and J.~B.~Wu,
  arXiv:0807.0802 .
  M.~A.~Bandres, A.~E.~Lipstein and J.~H.~Schwarz,
  arXiv:0807.0880 .
  Y.~Zhou,
  arXiv:0807.0890 .
  J.~Gomis, D.~Rodriguez-Gomez, M.~Van Raamsdonk and H.~Verlinde,
  arXiv:0807.1074 .
  T.~Li, Y.~Liu and D.~Xie,
  arXiv:0807.1183 .
  N.~Kim,
  arXiv:0807.1349 .
  A.~Hashimoto and P.~Ouyang,
  arXiv:0807.1500 .
  D.~Astolfi, V.~G.~M.~Puletti, G.~Grignani, T.~Harmark and M.~Orselli,
  arXiv:0807.1527 .
  C.~Ahn and R.~I.~Nepomechie,
  arXiv:0807.1924 .
  D.~Bak and S.~J.~Rey,
  arXiv:0807.2063 .
  K.~Hanaki and H.~Lin,
  arXiv:0807.2074 .
  Y.~Imamura and K.~Kimura,
  arXiv:0807.2144 .
  B.~H.~Lee, K.~L.~Panigrahi and C.~Park,
  arXiv:0807.2559 .
  I.~Shenderovich,
  arXiv:0807.2861 .
  C.~Ahn, P.~Bozhilov and R.~C.~Rashkov,
  arXiv:0807.3134 .
  Y.~Honma, S.~Iso, Y.~Sumitomo, H.~Umetsu and S.~Zhang,
  arXiv:0807.3825 .
  T.~McLoughlin and R.~Roiban,
  arXiv:0807.3965 .
  J.~Kluson,
  arXiv:0807.4054 .
  L.~F.~Alday, G.~Arutyunov and D.~Bykov,
  arXiv:0807.4400 .
  C.~Krishnan,
  arXiv:0807.4561 .
  N.~Gromov and V.~Mikhaylov,
  arXiv:0807.4897 .
  O.~Aharony, O.~Bergman and D.~L.~Jafferis,
  arXiv:0807.4924 .
  H.~Singh,
  arXiv:0807.5016 .
  G.~Bonelli, A.~Tanzini and M.~Zabzine,
  arXiv:0807.5113 .
\bibitem{Herzog:2002fn}
  C.~P.~Herzog,
  JHEP {\bf 0212}, 026 (2002)
  [arXiv:hep-th/0210126].
\bibitem{Klebanov:1996un}
  I.~R.~Klebanov and A.~A.~Tseytlin,
  Nucl.\ Phys.\  B {\bf 475}, 164 (1996)
  [arXiv:hep-th/9604089].
\bibitem{Garousi:2008ik}
  M.~R.~Garousi, A.~Ghodsi and M.~Khosravi,
  arXiv:0807.1478 [hep-th].
\bibitem{Son:2002sd}
  D.~T.~Son and A.~O.~Starinets,
  JHEP {\bf 0209}, 042 (2002)
  [arXiv:hep-th/0205051].
\bibitem{Herzog:2002pc}
  C.~P.~Herzog and D.~T.~Son,
  JHEP {\bf 0303}, 046 (2003)
  [arXiv:hep-th/0212072].
\bibitem{Russo:1997mk}
  J.~G.~Russo and A.~A.~Tseytlin,
  Nucl.\ Phys.\  B {\bf 508}, 245 (1997)
  [arXiv:hep-th/9707134].
\bibitem
{AAT}{A.A. Tseytlin, Nucl. Phys. B {\bf 276} (1987) 391.}
\bibitem
{AAT1}{A.A. Tseytlin, Phys. Lett. B {\bf 176} (1986) 92; R.R.
Metsaev, A.A. Tseytlin, Phys. Lett. B {\bf 185} (1987) 52.}

\bibitem{Kallosh:1998qs}
  R.~Kallosh and A.~Rajaraman,
  Phys.\ Rev.\  D {\bf 58}, 125003 (1998)
  [arXiv:hep-th/9805041].
\bibitem{Tseytlin:2000sf}
  A.~A.~Tseytlin,
  Nucl.\ Phys.\  B {\bf 584}, 233 (2000)
  [arXiv:hep-th/0005072].

 
\bibitem{Gubser:1998nz}
  S.~S.~Gubser, I.~R.~Klebanov and A.~A.~Tseytlin,
  Nucl.\ Phys.\  B {\bf 534}, 202 (1998)
  [arXiv:hep-th/9805156].




\bibitem{Buchel:2008sh}
  A.~Buchel,
  Nucl.\ Phys.\  B {\bf 803}, 166 (2008)
  [arXiv:0805.2683 [hep-th]].



\bibitem{Sen:2005wa}
  A.~Sen,
  JHEP {\bf 0509}, 038 (2005)
  [arXiv:hep-th/0506177].

\bibitem{Gross:1986iv}
  D.~J.~Gross and E.~Witten,
  Nucl.\ Phys.\  B {\bf 277}, 1 (1986).






\end{thebibliography}
\end{document}